\documentclass[preprint, superscriptaddress, showpacs,preprintnumbers,amsmath,amssymb,pra]{revtex4}
\usepackage{graphicx}

\begin{document}

\thispagestyle{empty}

\title{Nernst heat theorem for the thermal Casimir interaction
between two graphene sheets
}

\author{V.~B.~Bezerra}
\affiliation{Department of Physics, Federal University of Para\'{\i}ba,
C.P.5008, CEP 58059--970, Jo\~{a}o Pessoa, Pb-Brazil}
\author{
G.~L.~Klimchitskaya}

\affiliation{Central Astronomical Observatory
at Pulkovo of the Russian Academy of Sciences,
Saint Petersburg, 196140, Russia}
\affiliation{Institute of Physics, Nanotechnology and
Telecommunications, Peter the Great Saint Petersburg
Polytechnic University, Saint Petersburg, 195251, Russia}
\affiliation{Department of Physics, Federal University of Para\'{\i}ba,
C.P.5008, CEP 58059--970, Jo\~{a}o Pessoa, Pb-Brazil}

\author{
 V.~M.~Mostepanenko}
\affiliation{Central Astronomical Observatory
at Pulkovo of the Russian Academy of Sciences,
Saint Petersburg, 196140, Russia}
\affiliation{Institute of Physics, Nanotechnology and
Telecommunications, Peter the Great Saint Petersburg
Polytechnic University, Saint Petersburg, 195251, Russia}
\affiliation{Kazan Federal University, Kazan, 420008, Russia}
\affiliation{Department of Physics, Federal University of Para\'{\i}ba,
C.P.5008, CEP 58059--970, Jo\~{a}o Pessoa, Pb-Brazil}

\author{C.~Romero}
\affiliation{Department of Physics, Federal University of Para\'{\i}ba,
C.P.5008, CEP 58059--970, Jo\~{a}o Pessoa, Pb-Brazil}

\begin{abstract}
We find analytic asymptotic expressions at low temperature for
the Casimir free energy, entropy and pressure of two parallel
graphene sheets in the framework of the Lifshitz theory. The
reflection coefficients of electromagnetic waves on graphene
are described on the basis of first principles of quantum
electrodynamics at nonzero temperature using the polarization
tensor in (2+1)-dimensional space-time. The leading contributions
to the Casimir entropy and to the thermal corrections to the
Casimir energy and pressure are given by the thermal correction
to the polarization tensor at nonzero Matsubara frequencies.
It is shown that the Casimir entropy for two graphene sheets
goes to zero when the temperature vanishes, i.e., the third law
of thermodynamics (the Nernst heat theorem) is satisfied.
At low temperature, the magnitude of the thermal correction to
the Casimir pressure between two graphene sheets is shown to vary
inversely proportional to the separation. The Nernst heat theorem
for graphene is discussed in the context of problems occurring
in Casimir physics for both metallic and dielectric plates.
\end{abstract}
%\pacs{}

\maketitle

\section{Introduction}

The Casimir interaction arises between two closely spaced
material bodies due to fluctuations of the electromagnetic
field \cite{1}. At zero temperature, this interaction is
caused by the zero-point photons. At nonzero temperature,
there is one more contribution to the Casimir interaction
due to the thermal photons. Theoretical description of the
thermal Casimir force determined by both the zero-point and
thermal photons is given by the Lifshitz theory \cite{1,2}.
In the context of this theory, the free energy of the
Casimir interaction is presented as a functional of the
frequency-dependent dielectric permittivities of material
bodies.

During the last few years, the Casimir effect has attracted
considerable interest due to the important role it plays in
different fields of physics. In atomic physics, it determines
the atom-wall interaction and should be taken into account
in the phenomena of quantum reflection and Bose-Einstein
condensation \cite{3,4,5,6,7,8}.
In condensed matter physics, the Casimir effect was investigated
in connection with the role of temperature, conductivity
properties, surface roughness and phase transitions
\cite{9,10,11,12,13}.
There are also prospective applications of the Casimir effect
in nanotechnology to actuate micromechanical systems,
to explain stiction phenomena, to create new types of
microchips, etc. \cite{14,15,16,17,18}.

After the advent of graphene a lot of theoretical work has
been done to calculate the Casimir free energy and force
between two graphene sheets, between a graphene sheet and
a plate made of ordinary material, between two graphene-coated
substrates, and between an atom and a graphene sheet
\cite{19,20,21,22,23,24,25,26,27,28,29,30,31,32,33}.
It was found that, as opposed to ordinary materials, the
Casimir force between two graphene sheets has large thermal
contribution at short separations of about 100\,nm \cite{20}
(for ordinary materials the thermal contribution becomes dominant
at separations of about $6\,\mu$m \cite{1}). This is explained by
the fact that at low energies the quasiparticles in graphene
are massless Dirac fermions, which are described by the linear
dispersion relation, but move with the Fermi velocity $v_F$
rather than with the speed of light $c$ \cite{34}.
The major contribution to the thermal Casimir free energy and
force between two graphene sheets is given by the transverse
magnetic (TM), i.e., $p$-polarized, electromagnetic waves.
This is akin to the configuration of two metallic plates
described by the Drude model. In that case, there is a relatively
large thermal effect at short separations, and
the contribution of
the transverse electric (TE), i.e., $s$-polarized,
electromagnetic waves vanishes for plate separations above
$6\,\mu$m. The question arises  whether the Casimir free
energy and entropy of two graphene sheets meet the basic
requirements of thermodynamics and, specifically, the Nernst
heat theorem.

The point is that there is an outstanding unresolved problem
in Lifshitz theory \cite{1,35}. For metallic bodies made of
both nonmagnetic and magnetic metals with perfect crystal
lattices, it was shown \cite{36,37,38,39,40} that if the
low-frequency behavior of the dielectric permittivity of plate
metals is described by the dissipative Drude model, the Casimir
free energy and entropy calculated using the Lifshitz theory
violate the third law of thermodynamics known as the Nernst heat
theorem. In doing so, the Casimir entropy at zero temperature
takes a nonzero value which depends on the parameters of the
system. If some nonzero relaxation at zero temperature is
preserved due to the role of impurities, the Casimir entropy
abruptly jumps to zero at very low temperature \cite{41,42,42a},
so that the Nernst theorem is formally restored. This, however,
does not solve the problem because a perfect crystal lattice is
a truly equilibrium system with the nondegenerate state of lowest
energy, so that, according to quantum statistical physics, the
Nernst heat theorem must be applicable to it \cite{1,35}.
It was shown also that if a lossless plasma model is used
instead of the Drude one, the Lifshitz theory is found in
perfect agreement with the Nernst heat theorem
\cite{36,37,38,39,40}.

It was noted \cite{42b} that when the
spatially nonlocal dielectric response is considered, the Nernst
heat theorem is satisfied because the effects of spatial
dispersion lead to an effective residual relaxation even for a
perfect crystal lattice. At sufficiently short separations,
however, the frequency region of infrared optics, where the
dielectric response is local, plays a major role. As a result,
when the frequency regions with  both nonlocal and local response
functions are taken into account, the Nernst heat theorem is
again violated \cite{42c}.

As for experiment, all precise measurements of the Casimir
interaction between nonmagnetic (Au) and magnetic (Ni) test
 bodies exclude theoretical predictions of the Lifshitz theory
 using the Drude model and are in  good agreement with those
 using the plasma model \cite{43,44,44a,45,46,47,48}.
 Recently, differential Casimir experiments have been
 proposed \cite{49,50,51}, where the theoretical predictions
 calculated with the help of the Drude and plasma models differ
by up to a factor of 1000. According to the results of one of
these experiments, already performed, the Drude model is
unequivocally ruled out by the measurement data, whereas the
plasma model shows a good agreement with the data \cite{52,53}.
These results can be considered as surprising because in real
electromagnetic fields with a nonzero expectation value the
low-frequency response of metals is undoubtedly described by
the dissipative Drude model. This fact is confirmed by
thousands of different experiments. Thus, one can guess that
the response of metals to quantum fluctuations, having zero
expectation value, might be different.

For dielectric plates, the Lifshitz theory violates the
Nernst heat theorem if the dc conductivity of plate materials
is taken into account in calculations \cite{54,55,56,57,58}.
If the dc conductivity is disregarded, the Lifshitz theory
is proved to be in  perfect agreement with the Nernst heat
theorem \cite{54,55,56,57,58}. The experimental data of several
precise experiments on measuring the Casimir-Polder and
Casimir interaction with dielectric plates \cite{59,60,61,62}
were found in agreement with theory if the dc conductivity of
plate materials is  disregarded in calculations.
If, however, the dc conductivity of dielectric plates is taken
into account, the theoretical results are excluded by the
measurement data \cite{60,61,62,63}. This is again surprising
because at nonzero temperature the conductivity of dielectrics
at a constant current is, although small, a really existing and
measurable effect. We emphasize, however, that for both
dielectrics and metals the experimentally consistent calculations
are always found to be in agreement with the Nernst heat theorem.
Because of this, the latter can be considered as some kind of
 test for any novel theoretical approach.

In this paper, we derive the low-temperature asymptotic behavior
of the Casimir free energy, entropy and
pressure in the configuration of
two parallel graphene sheets. As already mentioned, the Casimir
 effect for two graphene sheets was investigated by many authors
and several unusual properties, as compared to metallic plates,
have been discovered. In the framework of the Dirac model,
the most fundamental description of the response of graphene to
electromagnetic field is given by the polarization tensor in
$(2+1)$-dimensional space-time. At zero temperature, the
polarization tensor of graphene was found in Ref.~\cite{19}.
At nonzero temperature it was derived in Ref.~\cite{24} at
the discrete Matsubara frequencies (the comparison with other
theoretical approaches to the Casimir effect in graphene systems
is contained in Refs.~\cite{29,30}). The Lifshitz theory with
the exact reflection coefficients expressed via the polarization
tensor \cite{19,24} has been used in many theoretical papers
\cite{27,28,29,30,31,32,64,65}. This formalism was applied in
the first experiment on measuring the Casimir interaction between
a Au-coated sphere and a graphene-coated substrate \cite{66}, and
a very good agreement with the measurement results was obtained
\cite{67}.

The derivation of our asymptotic expressions has been made
possible by the use of another representation for the
polarization tensor of graphene valid along the entire plane
of complex frequency \cite{68}. This representation was already
exploited to investigate the origin of large thermal effect in
graphene systems \cite{33}, the reflectivity properties of
graphene with a nonzero mass-gap parameter \cite{69} and
graphene-coated substrates \cite{70}, and the conductivity
properties of graphene \cite{71}.
According to our results, with vanishing temperature $T$ the
leading contributions to thermal correction to the  Casimir
energy and to the Casimir entropy
behave as $\sim T^3\ln{T}$ and $\sim T^2\ln{T}$, respectively.
This means that the Lifshitz theory for graphene is in
agreement with the Nernst heat theorem.

The paper is organized as follows. In Sec.~II, we present the
general formalism of the Lifshitz theory using the polarization
tensor of graphene at low temperature. In Sec.~III, the
thermal correction to the Casimir energy  is considered,
which arises due to a
summation over the discrete Matsubara frequencies when
the zero-temperature polarization tensor is used in calculations.
Section IV is devoted to the thermal correction arising due to
the temperature dependence of the polarization tensor. We
consider separately contributions from the zero-frequency term
and from all terms with nonzero Matsubara frequencies.
The total results for the low-temperature behavior of the
Casimir free energy, entropy and pressure for two graphene
sheets are presented in Sec.~V. Section VI contains our
conclusions and discussion. Some details of mathematical
derivations are given in Appendix A.

\section{General formalism in terms of the polarization tensor of
graphene at low temperature}

We consider the Casimir free energy per unit area of two
parallel graphene sheets separated by a gap of thickness $a$ at
temperature $T$ in thermal equilibrium with the environment
\cite{1,2}
\begin{equation}
{\cal F}(a,T)=\frac{k_BT}{2\pi}
\sum_{l=0}^{\infty}{\vphantom{\sum}}^{\prime}
\int_{0}^{\infty}k_{\bot}dk_{\bot}\left[
\ln\left(1-r_{\rm TM}^2e^{-2q_la}\right)\right.
+\left.\ln\left(1-r_{\rm TE}^2e^{-2q_la}\right)\right].
\label{eq1}
\end{equation}
\noindent
Here, $k_B$ is the Boltzmann constant, the prime on the
summation sign means that the term with $l=0$ is divided by 2,
$k_{\bot}$ is the magnitude of the projection of the wave
vector on the plane of plates, and
\begin{equation}
q_l=\sqrt{k_{\bot}^2+\frac{\xi_l^2}{c^2}},
\label{eq1a}
\end{equation}
\noindent
where $\xi_l=2\pi k_BTl/\hbar$ are the Matsubara frequencies.

The reflection coefficients $r_{\rm TM,TE}$ can be expressed
via the components of the polarization tensor of graphene,
$\Pi_{mn}$, with $m,n=0,\,1,\,2$ \cite{19,24,68}
\begin{eqnarray}
&&
r_{\rm TM}\equiv r_{\rm TM}(i\xi_l,k_{\bot})=
\frac{q_l\Pi_{00}(i\xi_l,k_{\bot})}{q_l\Pi_{00}(i\xi_l,k_{\bot})
+2\hbar k_{\bot}^2},
\nonumber \\
&&
r_{\rm TE}\equiv r_{\rm TE}(i\xi_l,k_{\bot})=
-\frac{\Pi(i\xi_l,k_{\bot})}{\Pi(i\xi_l,k_{\bot})
+2\hbar k_{\bot}^2q_l},
\label{eq2}
\end{eqnarray}
\noindent
where the combination of the components $\Pi$ is defined as
\begin{equation}
\Pi(i\xi_l,k_{\bot})=k_{\bot}^2\Pi_{\rm tr}(i\xi_l,k_{\bot})
-q_l^2\Pi_{00}(i\xi_l,k_{\bot})
\label{3}
\end{equation}
\noindent
and $\Pi_{\rm tr}\equiv\Pi_m^{\,m}$.

Note that the polarization tensor is directly connected with the
longitudinal and transverse nonlocal dielectric permittivities of
graphene \cite{25,30,72}
\begin{eqnarray}
&&
\varepsilon_{\|}(i\xi_l,k_{\bot})=1+\frac{1}{2\hbar k_{\bot}}
\Pi_{00}(i\xi_l,k_{\bot}),
\nonumber \\
&&
\varepsilon_{\bot}(i\xi_l,k_{\bot})=1+\frac{c^2}{2\hbar k_{\bot}\xi_l^2}
\Pi(i\xi_l,k_{\bot}).
\label{eq4}
\end{eqnarray}
\noindent
Thus, for graphene the dielectric permittivity is calculated
precisely starting from first principles of quantum field theory
at nonzero temperature.
This is different from more complicated materials which are
usually described by phenomenological dielectric functions.

For obtaining the low-temperature asymptotic expressions for the
Casimir free energy, it is convenient to use the representation
of the polarization tensor obtained in Ref.~\cite{68} and applied
in Refs.~\cite{33,69,70,71} (see also Ref.~\cite{73}, where this
representation was generalized for the case of nonzero chemical
potential). At first, the quantities $\Pi_{00}$ and $\Pi$ are
presented as the sums of zero-temperature contributions and the
thermal corrections to them
\begin{eqnarray}
&&
\Pi_{00}(i\xi,k_{\bot})=\Pi_{00}^{(0)}(i\xi,k_{\bot})
+\Delta_{T}\Pi_{00}(i\xi,k_{\bot}),
\nonumber \\
&&
\Pi(i\xi,k_{\bot})=\Pi^{(0)}(i\xi,k_{\bot})
+\Delta_{T}\Pi(i\xi,k_{\bot}).
\label{eq5}
\end{eqnarray}
\noindent
Note that for arbitrary $\xi$ the quantities $\Pi_{00}^{(0)}$ and
$\Pi^{(0)}$ are temperature independent. If, however, $\xi$ is
replaced with $\xi_l$, they depend on $T$ through the Matsubara
frequencies.

Below we consider pure (gapless) graphene sheets. For the
zero-temperature contributions in Eq.~(\ref{eq5}), calculated at
the Matsubara frequencies, one has \cite{19,24,68}
\begin{eqnarray}
&&
\Pi_{00}^{(0)}(i\xi_l,k_{\bot})=
\frac{\pi\alpha\hbar k_{\bot}^2}{\tilde{q}_l},
\nonumber \\
&&
\Pi^{(0)}(i\xi_l,k_{\bot})=
{\pi\alpha\hbar k_{\bot}^2}{\tilde{q}_l},
\label{eq6}
\end{eqnarray}
\noindent
where $\alpha=e^2/(\hbar c)$ is the fine structure constant,
\begin{equation}
\tilde{q}_l=\sqrt{\tilde{v}_F^2k_{\bot}^2+\frac{\xi_l^2}{c^2}},
\label{eq7}
\end{equation}
\noindent
and $\tilde{v}_F=v_F/c\approx 1/300$.

The respective zero-temperature reflection coefficients are
obtained by the substitution of Eq.~(\ref{eq6}) in
Eq.~(\ref{eq2}).
They have the following form:
\begin{eqnarray}
&&
r_{\rm TM}^{(0)}\equiv
r_{\rm TM}^{(0)}(i\xi_l,k_{\bot})=
\frac{q_l\Pi_{00}^{(0)}(i\xi_l,k_{\bot})}{q_l
\Pi_{00}^{(0)}(i\xi_l,k_{\bot})+2\hbar k_{\bot}^2}
=\frac{\alpha\pi q_l}{\alpha\pi q_l+2\tilde{q}_l},
\nonumber \\
&&\label{eq8}\\[-2mm]
&&
r_{\rm TE}^{(0)}\equiv
r_{\rm TE}^{(0)}(i\xi_l,k_{\bot})=
-\frac{\Pi^{(0)}(i\xi_l,k_{\bot})}{\Pi^{(0)}(i\xi_l,k_{\bot})
+2\hbar k_{\bot}^2q_l}
=
-\frac{\alpha\pi\tilde{q}_l}{\alpha\pi\tilde{q}_l+2q_l}.
\nonumber
\end{eqnarray}

The thermal corrections on the right-hand side of Eq.~(\ref{eq5})
at the Matsubara frequencies are most conveniently represented
in Eq.~(10) of Ref.~\cite{33}
\begin{eqnarray}
&&
\Delta_T\Pi_{00}(i\xi_l,k_{\bot})=
\frac{8\alpha\hbar\tilde{q}_l}{\tilde{v}_F^2}
\int_{0}^{\infty}\frac{du}{e^{B_lu}+1}
\nonumber \\
&&~~
\times\left\{1-\frac{1}{\sqrt{2}}\left[
\sqrt{(1+u^2)^2-4\frac{\tilde{v}_F^2k_{\bot}^2u^2}{\tilde{q}_l^2}}
+1-u^2\right]^{1/2}\right\},
\nonumber \\[1mm]
&&
\Delta_T\Pi(i\xi_l,k_{\bot})=
\frac{8\alpha\hbar\tilde{q}_l}{\tilde{v}_F^2}
\int_{0}^{\infty}\frac{du}{e^{B_lu}+1}
\nonumber \\
&&~~
\times\left\{
\vphantom{\left[\frac{\tilde{v}_F^2k_{\bot}^2}{\tilde{q}_l^2
\sqrt{(1+u^2)^2-
4\frac{\tilde{v}_F^2k_{\bot}^2u^2}{\tilde{q}_l^2}}}\right]}
-\frac{\xi_l^2}{c^2}+\frac{\tilde{q}_l^2}{\sqrt{2}}\left[
\sqrt{(1+u^2)^2-4\frac{\tilde{v}_F^2k_{\bot}^2u^2}{\tilde{q}_l^2}}
+1-u^2\right]^{1/2}\right.
\nonumber \\
&&~~\left.
\times\left[1-\frac{\tilde{v}_F^2k_{\bot}^2}{\tilde{q}_l^2
\sqrt{(1+u^2)^2-
4\frac{\tilde{v}_F^2k_{\bot}^2u^2}{\tilde{q}_l^2}}}\right]
\right\},
\label{eq9}
\end{eqnarray}
where $B_l\equiv\hbar c\tilde{q}_l/(2k_BT)$.

As is seen from Eq.~(\ref{eq9}),
\begin{equation}
{\displaystyle{\lim_{T\to 0}}}\Delta_T\Pi_{00}(i\xi_l,k_{\bot})=
{\displaystyle{\lim_{T\to 0}}}\Delta_T\Pi(i\xi_l,k_{\bot})=0,
\label{eq10}
\end{equation}
\noindent
whereas for the zero-temperature contributions calculated
at the Matsubara frequencies one has:
\begin{eqnarray}
&&
{\displaystyle{\lim_{T\to 0}}}\Pi_{00}^{(0)}(i\xi_l,k_{\bot})=
\frac{\pi\alpha\hbar k_{\bot}}{\tilde{v}_F}\neq 0,
\nonumber \\
&&
{\displaystyle{\lim_{T\to 0}}}\Pi^{(0)}(i\xi_l,k_{\bot})=
\pi\alpha\hbar\tilde{v}_F k_{\bot}^3\neq 0.
\label{eq11}
\end{eqnarray}

Thus, for sufficiently low $T$, one can use two small parameters,
namely,
\begin{equation}
\frac{\Delta_T\Pi_{00}}{\Pi_{00}^{(0)}}\ll 1, \quad
\frac{\Delta_T\Pi}{\Pi^{(0)}}\ll 1.
\label{eq12}
\end{equation}
\noindent
An explicit condition on the smallness of $T$, wherein the
inequalities (\ref{eq12}) are satisfied, is determined in
Secs.~III and IV.

Now, we substitute Eq.~(\ref{eq5}) in  Eq.~(\ref{eq2}), expand in
powers of the small parameters (\ref{eq12}) and preserve only the
first-order terms. Then, using  Eq.~(\ref{eq8}), we obtain
\begin{eqnarray}
&&
r_{\rm TM}=r_{\rm TM}^{(0)}+
\frac{2\hbar k_{\bot}^2r_{\rm TM}^{(0)}}{q_l\Pi_{00}^{(0)}+
2\hbar k_{\bot}^2}\,\frac{\Delta_T\Pi_{00}}{\Pi_{00}^{(0)}},
\nonumber \\
&&
r_{\rm TE}=r_{\rm TE}^{(0)}+
\frac{2\hbar k_{\bot}^2q_lr_{\rm TE}^{(0)}}{\Pi^{(0)}+
2\hbar k_{\bot}^2q_l}\,\frac{\Delta_T\Pi}{\Pi^{(0)}}.
\label{eq13}
\end{eqnarray}
\noindent
Taking the square of Eq.~(\ref{eq13}) and using Eq.~(\ref{eq6}),
in the first order of the small parameters (\ref{eq12}) we find
\begin{eqnarray}
&&
r_{\rm TM}^2={r_{\rm TM}^{(0)}\!}^2\left(1+
\frac{4\tilde{q}_l}{\pi\alpha q_l+2\tilde{q}_l}
\,\frac{\Delta_T\Pi_{00}}{\Pi_{00}^{(0)}}\right),
\nonumber \\
&&
r_{\rm TE}^2={r_{\rm TE}^{(0)}}^2\left(1+
\frac{4q_l}{\pi\alpha\tilde{q}_l+2q_l}\,
\frac{\Delta_T\Pi}{\Pi^{(0)}}\right).
\label{eq14}
\end{eqnarray}

Substituting Eq.~(\ref{eq14}) in Eq.~(\ref{eq1}) and expanding
the logarithms up to the first powers of small parameters
(\ref{eq12}), one arrives at
\begin{eqnarray}
&&
{\cal F}(a,T)\equiv{\cal F}^{(1)}(a,T)+\Delta_T^{\!(2)}{\cal F}(a,T)
\nonumber \\
&&~~
=\frac{k_BT}{2\pi}
\sum_{l=0}^{\infty}{\vphantom{\sum}}^{\prime}
\int_{0}^{\infty}k_{\bot}dk_{\bot}\left[
\ln\left(1-{r_{\rm TM}^{(0)}\!}^2e^{-2q_la}\right)\right.
+\left.\ln\left(1-{r_{\rm TE}^{(0)}\!}^2e^{-2q_la}\right)\right]
\label{eq15}
\\
&&
-\frac{2k_BT}{\pi}
\sum_{l=0}^{\infty}{\vphantom{\sum}}^{\prime}\!
\int_{0}^{\infty}\!\!\!\!k_{\bot}dk_{\bot}\!\left[
\frac{\tilde{q}_l}{\pi\alpha q_l+2\tilde{q}_l}
\frac{{r_{\rm TM}^{(0)}\!}^2}{e^{2q_la}-{r_{\rm TM}^{(0)}\!}^2}
\frac{\Delta_T\Pi_{00}}{\Pi_{00}^{(0)}}\right.
\nonumber\\
&&~~~~~
\left.
+\frac{{q}_l}{\pi\alpha \tilde{q}_l+2{q}_l}
\frac{{r_{\rm TE}^{(0)}\!}^2}{e^{2q_la}-{r_{\rm TE}^{(0)}\!}^2}
\frac{\Delta_T\Pi}{\Pi^{(0)}}\right],
\nonumber
\end{eqnarray}
\noindent
where ${\cal F}^{(1)}$ and $\Delta_T^{\!(2)}{\cal F}$
are equal to the first
and second primed sums, respectively.

Now it is essential to have in mind the Lifshitz formula for the
Casimir energy per unit area of graphene sheet at zero
temperature,
which reads as
 \cite{1}
\begin{equation}
{E}(a)=\frac{\hbar}{4\pi^2}
\int_{0}^{\infty}d\xi
\int_{0}^{\infty}k_{\bot}dk_{\bot}\left[
\ln\left(1-{r_{\rm TM}^{(0)}\!}^2e^{-2qa}\right)\right.
+\left.\ln\left(1-{r_{\rm TE}^{(0)}\!}^2e^{-2qa}\right)\right],
\label{eq16}
\end{equation}
\noindent
where
\begin{equation}
r_{\rm TM,TE}^{(0)}\equiv  r_{\rm TM,TE}^{(0)}(i\xi,k_{\bot}),
\quad
q=\sqrt{k_{\bot}^2+\frac{\xi^2}{c^2}}
\label{eq17}
\end{equation}
\noindent
are functions of the continuous variable $\xi$.

It is seen that the first sum in Eq.~(\ref{eq15}) is
obtained from Eq.~(\ref{eq16}) by the familiar substitution
\begin{equation}
\frac{\hbar}{4\pi^2}
\int_{0}^{\infty}d\xi \to
k_BT\sum_{l=0}^{\infty}{\vphantom{\sum}}^{\prime}.
\label{eq18}
\end{equation}
\noindent
Because of this, it is convenient to represent this sum
in the form
\begin{equation}
{\cal F}^{(1)}(a,T)=E(a)+\Delta_T^{\!(1)}{\cal F}(a,T),
\label{eq19}
\end{equation}
\noindent
where $\Delta_T^{\!(1)}{\cal F}(a,T)$ has the meaning of the
first part of the thermal correction to the Casimir energy
of two graphene  sheets. The thermal correction
$\Delta_T^{\!(1)}{\cal F}(a,T)$ is determined by the fact that
at nonzero $T$ the continuous argument $\xi$ in the
zero-temperature polarization tensor $\Pi_{mn}^{(0)}$ and the
respective reflection coefficients $r_{\rm TM,TE}^{(0)}$ is
replaced by the discrete Matsubara frequencies $\xi_l$, and
originates from a summation over these frequencies.

The second sum in Eq.~(\ref{eq15}), which we have notated
$\Delta_T^{\!(2)}{\cal F}(a,T)$, has the meaning of the
second part of thermal correction to the Casimir energy
of two graphene  sheets.  It originates from an explicit
dependence of the polarization tensor $\Pi_{mn}$ on the
temperature as a parameter. With account of Eq.~(\ref{eq19}),
Eq.~(\ref{eq15}) can be rewritten as
\begin{equation}
{\cal F}(a,T)=E(a)+\Delta_T^{\!(1)}{\cal F}(a,T)
+\Delta_T^{\!(2)}{\cal F}(a,T).
\label{eq20}
\end{equation}

Below it is convenient to determine the individual asymptotic
behaviors of each part of the total thermal correction
\begin{equation}
\Delta_T{\cal F}(a,T)=\Delta_T^{\!(1)}{\cal F}(a,T)
+\Delta_T^{\!(2)}{\cal F}(a,T)
\label{eq21}
\end{equation}
\noindent
at low temperature separately. This is done in the next two sections.

\section{Thermal correction to the Casimir energy due to
{\protect \\}summation over the Matsubara frequencies}

In this section, we investigate the low-temperature behavior of
the thermal correction $\Delta_T^{\!(1)}{\cal F}$ defined
in Eq.~(\ref{eq19}). It is convenient to represent the
quantity ${\cal F}^{(1)}(a,T)$ in terms of the dimensionless
variables
\begin{equation}
y=2aq_l,\quad \zeta_l=\frac{\xi_l}{\omega_c}\equiv
\frac{2a\xi_l}{c}\equiv\tau l,
\label{eq22}
\end{equation}
\noindent
where
\begin{equation}
\tau=4\pi\frac{ak_BT}{\hbar c}=2\pi\frac{T}{T_{\rm eff}},
\quad
k_BT_{\rm eff}\equiv\hbar\omega_c.
\label{eq23}
\end{equation}
\noindent
Then, using Eq.~(\ref{eq15}), one arrives at
\begin{equation}
{\cal F}^{(1)}=\frac{k_BT}{8\pi a^2}
\sum_{l=0}^{\infty}{\vphantom{\sum}}^{\prime}
\Phi(\tau l),
\label{eq24}
\end{equation}
\noindent
where
\begin{equation}
\Phi(\tau l)=\int_{\tau l}^{\infty\!\!}ydy
\left[\ln\left(1-{r_{\rm TM}^{(0)}}^2e^{-y}\right)+
\ln\left(1-{r_{\rm TE}^{(0)}}^2e^{-y}\right)\right]
\label{eq25}
\end{equation}
\noindent
and the reflection coefficients (\ref{eq8}) take the form
\begin{equation}
r_{\rm TM}^{(0)}=\frac{\alpha\pi y}{\alpha\pi y+2\tilde{g}_l},
\quad
r_{\rm TE}^{(0)}=
\frac{\alpha\pi\tilde{g}_l}{\alpha\pi\tilde{g}_l+2y}.
\label{eq26}
\end{equation}
\noindent
Here, the dimensionless function $\tilde{g}_l$ is defined as
\begin{equation}
\tilde{g}_l=2a\tilde{q}_l=
\sqrt{\tilde{v}_F^2y^2+(1-\tilde{v}_F^2)(\tau l)^2}
\approx\sqrt{\tilde{v}_F^2y^2+(\tau l)^2},
\label{eq27}
\end{equation}
\noindent
where we have neglected the small quantity $\tilde{v}_F^2$
as compared to unity.
Now we represent the Casimir energy at zero temperature, $E(a)$, in
terms of dimensionless variables
\begin{equation}
y=2aq,\quad \zeta=\frac{\xi}{\omega_c}\equiv\tau l,
\label{eq28}
\end{equation}
\noindent
where $q$ is defined in Eq.~(\ref{eq17}) and $\tau$ in
Eq.~(\ref{eq23}).
Then, from Eq.~(\ref{eq16}) one obtains
\begin{equation}
E(a)=\frac{k_BT}{8\pi a^2}\int_{0}^{\infty}\!\!dt
\Phi(\tau t).
\label{eq29}
\end{equation}
\noindent
Here, the function $\Phi$ is defined in Eqs.~(\ref{eq25}) and
(\ref{eq26}), where the discrete quantity $l$ is replaced with
the continuous variable $t$.

{}From the comparison of Eqs.~(\ref{eq24}) and (\ref{eq29}) it
is seen that the Casimir energy has the same form as the
quantity ${\cal F}^{(1)}$, but is represented by an integral
instead of a discrete sum. Then, the thermal correction
$\Delta_T^{\!(1)}{\cal F}(a,T)$, defined in Eq.~(\ref{eq19}),
can be found as the difference between the sum and the integral
\begin{equation}
\Delta_T^{\!(1)}{\cal F}(a,T)=\frac{k_BT}{8\pi a^2}
\left[\sum_{l=0}^{\infty}{\vphantom{\sum}}^{\prime}
\Phi(\tau l)-\int_{0}^{\infty}\!\!dt\Phi(\tau t)
\right].
\label{eq30}
\end{equation}

Using the Abel-Plana formula \cite{1}, we can rewrite
Eq.~(\ref{eq30}) in the form
\begin{equation}
\Delta_T^{\!(1)}{\cal F}(a,T)=\frac{ik_BT}{8\pi a^2}
\int_{0}^{\infty}\!\!dt
\frac{\Phi(i\tau t)-\Phi(-i\tau t)}{e^{2\pi t}-1}
.
\label{eq31}
\end{equation}

The major contribution to this integral is given by
$t\sim 1/(2\pi)$. At the same time, the major contribution
to the integral (\ref{eq25}) is given by $y\sim 1$.
Below, we consider sufficiently low $T$ such that for
$t\sim 1/(2\pi)$ and $y\sim 1$ one would have
$\tau t\ll\tilde{v}_Fy$, i.e., in accordance with
Eq.~(\ref{eq23}),
\begin{equation}
\frac{\tau}{2\pi}=\frac{T}{T_{\rm eff}}\ll\tilde{v}_F.
\label{eq32}
\end{equation}
\noindent
This inequality can be rewritten in the form
\begin{equation}
k_BT\ll\frac{\hbar{v}_F}{2a}\equiv k_BT_{\rm eff}^{(g)}.
\label{eq33}
\end{equation}
\noindent
Here, $T_{\rm eff}^{(g)}$ is the effective temperature for
graphene, where, as compared to the standard definition
(\ref{eq23}), the speed of light $c$ is replaced with the
Fermi velocity $v_F$ \cite{20}. Note that the region of $T$
defined in Eqs.~(\ref{eq32}) and (\ref{eq33}) depends on the
separation distance between two graphene sheets.
For example, for $a=10$ and 100\,nm
Eqs.~(\ref{eq32}) and (\ref{eq33}) lead to $T\ll 300\,$K
and $T\ll 30\,$K, respectively.

Expanding the logarithms in the definition of $\Phi$ in
a power series, one obtains
\begin{eqnarray}
&&
\Phi(\tau t)\equiv \Phi_{\rm TM}(\tau t)+
\Phi_{\rm TE}(\tau t)
\label{eq34} \\
&&
=-\sum_{n=1}^{\infty}\frac{1}{n}\!\int_{\tau t}^{\infty}\!\!
\!ydy\left[{r_{\rm TM}^{(0)}}^{2n}\!(i\tau t,y)+
{r_{\rm TE}^{(0)}}^{2n}\!(i\tau t,y)
\right]e^{-ny},
\nonumber
\end{eqnarray}
\noindent
where $\Phi_{\rm TM}$ and $\Phi_{\rm TE}$ are defined via
the respective reflection coeffecients $r_{\rm TM}^{(0)}$
and $r_{\rm TE}^{(0)}$.

Now we expand the powers of the reflection coefficients
(\ref{eq26}), appearing in  Eq.~(\ref{eq34}), in powers of the
small parameter $\tau t/(\tilde{v}_Fy)$ by preserving only
the lowest order contribution
\begin{eqnarray}
&&
{r_{\rm TM}^{(0)}\!}^{2n}(i\tau t,y)=\rho_{\rm TM}^{2n}
\left[1-2n\frac{\tilde{v}_F}{\alpha\pi+2\tilde{v}_F}
\left(\frac{\tau t}{\tilde{v}_Fy}\right)^2\right],
\nonumber \\[-0.5mm]
&&\label{eq35}\\[-0.5mm]
&&
{r_{\rm TE}^{(0)}\!}^{2n}(i\tau t,y)=\rho_{\rm TE}^{2n}
\left[1+2n\frac{1}{2+\alpha\pi\tilde{v}_F}
\left(\frac{\tau t}{\tilde{v}_Fy}\right)^2\right].
\nonumber
\end{eqnarray}
\noindent
The quantities $\rho_{\rm TM,TE}$, introduced here,
are defined as
\begin{eqnarray}
&&
\rho_{\rm TM}=\frac{\alpha\pi}{\alpha\pi+2\tilde{v}_F},
\quad \rho_{\rm TM}^2\approx 0.6,
\label{eq36} \\
&&
\rho_{\rm TE}=\frac{\alpha\pi\tilde{v}_F}{2+\alpha\pi\tilde{v}_F}
\approx\frac{\alpha\pi\tilde{v}_F}{2},
\quad \rho_{\rm TE}^2\approx 1.4\times 10^{-9}.
\nonumber
\end{eqnarray}
\noindent
Note that these coefficients coincide with those in
Eq.~(\ref{eq26}) for $l=0$.

Substituting Eqs.~(\ref{eq35}) and (\ref{eq36}) in
Eq.~(\ref{eq34}), we arrive at
\begin{eqnarray}
&&
\Phi_{\rm TM}(\tau t)=-\sum_{n=1}^{\infty}
\frac{\rho_{\rm TM}^{2n}}{n}\left[
\vphantom{\frac{2n(\tau t)^2}{\tilde{v}_F(\alpha\pi+2\tilde{v}_F)}}
R_n^{(1)}(\tau t)\right.
 -\left.
\frac{2n(\tau t)^2}{\tilde{v}_F(\alpha\pi+2\tilde{v}_F)}
R_n^{(2)}(\tau t)\right],
\nonumber \\[-0.5mm]
&&\label{eq37}\\[-0.5mm]
&&
\Phi_{\rm TE}(\tau t)=-\rho_{\rm TE}^{2}\left[R_1^{(1)}(\tau t) +
\frac{(\tau t)^2}{\tilde{v}_F^2}R_1^{(2)}(\tau t)\right],
\nonumber
\end{eqnarray}
\noindent
where the functions $R_n^{(1)}$ and $R_n^{(2)}$ are defined as
\begin{eqnarray}
&&
R_n^{(1)}(\tau t)\equiv\int_{\tau t}^{\infty}ydye^{-ny}
=\frac{1}{n^2}e^{-n\tau t}(1+n\tau t),
\nonumber \\
&&
R_n^{(2)}(\tau t)\equiv\int_{\tau t}^{\infty}
\frac{dy}{y}e^{-ny}=-{\rm Ei}(-n\tau t),
\label{eq38}
\end{eqnarray}
\noindent
and ${\rm Ei}(z)$ is the integral exponent function \cite{74}.
Note that in the second line of Eq.~(\ref{eq37}) we keep
only the first term in the sum with respect to $n$.
This is justified by the smallness of the quantity
 $\rho_{\rm TE}^2$,
in accordance to Eq.~(\ref{eq36}).

Expanding the right-hand sides
in Eq.~(\ref{eq38}) in powers of $n\tau t$ one obtains
\begin{eqnarray}
&&
R_n^{(1)}(\tau t)=\frac{1}{n^2}\left[1-\frac{n^2}{2}(\tau t)^2+
\frac{n^3}{3}(\tau t)^3+\ldots\,\right],
\nonumber \\
&&
R_n^{(2)}(\tau t)=-\left[C+\ln(n\tau t)-n\tau t+\ldots\,\right],
\label{eq39}
\end{eqnarray}
\noindent
where $C$ is the Euler constant.

Now we are in the position to calculate the leading term in the
difference $\Phi(i\tau t)-\Phi(-i\tau t)$ appearing in
Eq.~(\ref{eq31}).
This is done separately for $\Phi_{\rm TM}$ and $\Phi_{\rm TE}$.
At first, we substitute Eq.~(\ref{eq39}) in the first line of
Eq.~(\ref{eq37}) and to the leading order in $\tau t$ we find
\begin{eqnarray}
&&
\Phi_{\rm TM}(i\tau t)-\Phi_{\rm TM}(-i\tau t)=
\frac{2\tau^2t^2}{\tilde{v}_F(\alpha\pi +2\tilde{v}_F)}
\sum_{n=1}^{\infty}\rho_{\rm TM}^{2n}\left[\ln i-
\ln(-i)\right]
\nonumber \\
&&~~~~~~~~
=i\frac{2\pi\tau^2t^2}{\tilde{v}_F(\alpha\pi +2\tilde{v}_F)}
\frac{\rho_{\rm TM}^2}{1-\rho_{\rm TM}^2}.
\label{eq40}
\end{eqnarray}

Note that $R_n^{(1)}$ contributes to Eq.~(\ref{eq40})
only starting from the third-order term in Eq.~(\ref{eq39}),
i.e., would lead to a higher-order correction $\sim\tau^3$.
In a similar way, the next after the logarithm, linear, term in
$R_n^{(2)}$ would also lead to a correction of order
$\tau^3$ and, thus, both these corrections can be omitted.

Likewise, substituting Eq.~(\ref{eq39}) in the second line of
Eq.~(\ref{eq37}) and omitting the terms of order of $\tau^3$
and higher, one finds
\begin{equation}
\Phi_{\rm TE}(i\tau t)-\Phi_{\rm TE}(-i\tau t)=
-i\frac{\pi\tau^2t^2}{\tilde{v}_F^2}
\rho_{\rm TE}^2.
\label{eq41}
\end{equation}

Using Eq.~(\ref{eq36}), Eqs.~(\ref{eq40}) and (\ref{eq41}) can
be rewritten in the form
\begin{eqnarray}
&&
\Phi_{\rm TM}(i\tau t)-\Phi_{\rm TM}(-i\tau t)=
i\frac{\pi^3\alpha^2\tau^2t^2}{2\tilde{v}_F^2
(\alpha\pi +2\tilde{v}_F)(\alpha\pi +\tilde{v}_F)},
\nonumber \\
&&
\Phi_{\rm TE}(i\tau t)-\Phi_{\rm TE}(-i\tau t)=
-i\frac{1}{4}\pi^3\alpha^2\tau^2t^2.
\label{eq42}
\end{eqnarray}

Substituting these results in Eq.~(\ref{eq31}), for the TM and
TE contributions to the thermal correction
$\Delta_T^{\!(1)}{\cal F}$, we obtain
\begin{eqnarray}
&&
\Delta_T^{\!(1)}{\cal F}_{\rm TM}(a,T)=-\frac{k_BT}{16a^2}
\frac{\pi^2\alpha^2\tau^2}{\tilde{v}_F^2
(\alpha\pi +2\tilde{v}_F)(\alpha\pi +\tilde{v}_F)}
\int_{0}^{\infty}\!\!dt\frac{t^2}{e^{2\pi t}-1},
\label{eq43} \\
&&
\Delta_T^{\!(1)}{\cal F}_{\rm TE}(a,T)=\frac{k_BT}{32a^2}
\pi^2\alpha^2\tau^2
\int_{0}^{\infty}\!\!dt\frac{t^2}{e^{2\pi t}-1}.
\nonumber
\end{eqnarray}
\noindent
Calculating the integrals in this equation and returning
to the dimensional temperature $T$ in all factors, one
arrives at
\begin{eqnarray}
&&
\Delta_T^{\!(1)}{\cal F}_{\rm TM}(a,T)=-\frac{(k_BT)^3}{(\hbar v_F)^2}
\frac{\zeta(3)}{4}
\frac{\pi\alpha^2}{(\alpha\pi +2\tilde{v}_F)
(\alpha\pi +\tilde{v}_F)},
\nonumber \\
&&
\Delta_T^{\!(1)}{\cal F}_{\rm TE}(a,T)=
\frac{(k_BT)^3}{(\hbar c)^2}
\frac{\zeta(3)}{8}
\pi\alpha^2,
\label{eq44}
\end{eqnarray}
\noindent
where $\zeta(z)$ is the Riemann zeta function.

As is seen in Eq.~(\ref{eq44}), the contributions of the TM and
TE modes to the thermal correction $\Delta_T^{\!(1)}{\cal F}$ are
of the same order in temperature, but of opposite signs.
It is seen also that the TE contribution is negligibly small,
as compared to the magnitude of the TM contribution
\begin{equation}
\frac{\Delta_T^{\!(1)}{\cal F}_{\rm TE}}{|\Delta_T^{\!(1)}
{\cal F}_{\rm TM}|}=\frac{1}{2}\tilde{v}_F^2
(\alpha\pi +2\tilde{v}_F)(\alpha\pi +\tilde{v}_F)\ll 1.
\label{eq45}
\end{equation}

According to Eq.~(\ref{eq44}), both  contributions to
the thermal correction $\Delta_T^{\!(1)}{\cal F}$ do not depend
on the separation and, thus, do not contribute to the Casimir
pressure between two graphene sheets (see Sec.~V for further
discussion).

At the end of this section, we underline that the higher-order
terms in the small parameter $\tau t/(\tilde{v}_Fy)$ in
Eq.~(\ref{eq35}), neglected in our calculation, result in
corrections of order $\tau^3t^3$ and higher in the difference
$\Phi(i\tau t)-\Phi(-i\tau t)$. These corrections lead to
terms of order $T^4$, which can be neglected in comparison with
the leading terms found in Eq.~(\ref{eq44}).
Note that several applications of the Abel-Plana formula for
determination of the low-temperature behavior of the Casimir
force between different materials are considered in
Ref.~\cite{78a}.

\section{Thermal correction to the Casimir energy due to
{\protect \\}temperature dependence of the polarization tensor}

In this section, we investigate the asymptotic behavior at
low $T$ of the second part of the thermal correction to the Casimir
energy between two graphene sheets,
$\Delta_T^{\!(2)}{\cal F}(a,T)$, defined by the second sum in
$l$ in Eq.~(\ref{eq15}). It is convenient to consider separately
contributions to $\Delta_T^{\!(2)}{\cal F}$
of the zero-frequency term  and of all terms with nonzero
Matsubara frequencies.

\subsection{Contribution of the zero-frequency term}

According to Eq.~(\ref{eq15}), the zero-frequency contribution to
the second part of the thermal correction
$\Delta_T^{\!(2)}{\cal F}$
is represented by the sum of TM and TE modes
\begin{equation}
\Delta_T^{\!(2)}{\cal F}^{(l=0)}(a,T)=
\Delta_T^{\!(2)}{\cal F}_{\rm TM}^{(l=0)}(a,T)+
\Delta_T^{\!(2)}{\cal F}_{\rm TE}^{(l=0)}(a,T),
\label{eq46}
\end{equation}
where
\begin{eqnarray}
&&
\Delta_T^{\!(2)}{\cal F}_{\rm TM}^{(l=0)}(a,T)=-\frac{k_BT}{\pi}
\frac{\tilde{v}_F}{\pi\alpha+2\tilde{v}_F}
\int_{0}^{\infty}\!\!k_{\bot}dk_{\bot}
\frac{{r_{\rm TM}^{(0)}}^2}{e^{2k_{\bot}a}-{r_{\rm TM}^{(0)}}^2}
\frac{\Delta_T\Pi_{00}(0,k_{\bot})}{\Pi_{00}^{(0)}(0,k_{\bot})},
\nonumber\\
&&
\Delta_T^{\!(2)}{\cal F}_{\rm TE}^{(l=0)}(a,T)=-\frac{k_BT}{\pi}
\frac{1}{\pi\alpha\tilde{v}_F+2}
\int_{0}^{\infty}\!\!k_{\bot}dk_{\bot}
\frac{{r_{\rm TE}^{(0)}}^2}{e^{2k_{\bot}a}-{r_{\rm TE}^{(0)}}^2}
\frac{\Delta_T\Pi(0,k_{\bot})}{\Pi^{(0)}(0,k_{\bot})}.
\label{eq47}
\end{eqnarray}

The thermal corrections $\Delta_T\Pi_{00}(0,k_{\bot})$ and
$\Delta_T\Pi(0,k_{\bot})$ are obtained from Eq.~(\ref{eq9}) by
putting $l=0$
\begin{eqnarray}
&&
\Delta_T\Pi_{00}(0,k_{\bot})=
\frac{8\alpha\hbar k_{\bot}}{\tilde{v}_F}\left(
\int_{0}^{\infty}\frac{du}{e^{B_0u}+1}\right.
-\left.
\int_{0}^{1}\frac{du}{e^{B_0u}+1}\sqrt{1-u^2}\right),
\label{eq48} \\
&&
\Delta_T\Pi(0,k_{\bot})=-8\alpha\hbar\tilde{v}_Fk_{\bot}^3
\int_{0}^{1}\frac{du}{e^{B_0u}+1}
\frac{u^2}{\sqrt{1-u^2}},
\nonumber
\end{eqnarray}
\noindent
where $B_0=\hbar c\tilde{v}_Fk_{\bot}/(2k_BT)$.
These expressions coincide up to a change of the integration
variable with Eq.~(51) of Ref.~\cite{68}. As shown in
Ref.~\cite{68}, by means of identical transformations the
expressions (\ref{eq48}) can be put in an equivalent form
\begin{eqnarray}
&&
\Delta_T\Pi_{00}(0,k_{\bot})=
\frac{16\alpha k_BT}{\tilde{v}_F^2c}
\int_{0}^{1}dx\ln\left[1+e^{-2B_0\sqrt{x(1-x)}}\right],
\nonumber \\
&&
\Delta_T\Pi(0,k_{\bot})=-16\alpha\hbar\tilde{v}_Fk_{\bot}^3
\int_{0}^{1}\frac{\sqrt{x(1-x)}dx}{e^{2B_0\sqrt{x(1-x)}}+1},
\label{eq49}
\end{eqnarray}
\noindent
which is convenient for us now.

Substituting Eq.~(\ref{eq49}) in Eq.~(\ref{eq47}) and using
Eq.~(\ref{eq6}) at $\xi_0=0$, we find
\begin{eqnarray}
&&
\Delta_T^{\!(2)}{\cal F}_{\rm TM}^{(l=0)}(a,T)=
-\frac{16(k_BT)^2}{\pi^2\hbar c(\pi\alpha+2\tilde{v}_F)}
\int_{0}^{\infty}\!\!\!dk_{\bot}
\frac{\rho_{\rm TM}^2}{e^{2k_{\bot}a}-\rho_{\rm TM}^2}
\int_{0}^{1\!\!\!}dx\ln\left[1+e^{-2B_0\sqrt{x(1-x)}}\right],
\nonumber\\
&&
\Delta_T^{\!(2)}{\cal F}_{\rm TE}^{(l=0)}(a,T)=
\frac{32k_BT}{\pi^2(\pi\alpha\tilde{v}_F+2)}
\int_{0}^{\infty}\!\!\!k_{\bot}dk_{\bot}
\frac{\rho_{\rm TE}^2}{e^{2k_{\bot}a}-\rho_{\rm TE}^2}
\int_{0}^{1}\!\!\frac{\sqrt{x(1-x)}dx}{e^{2B_0\sqrt{x(1-x)}}+1}.
\label{eq50}
\end{eqnarray}

We consider first the contribution of the TM mode. For this
purpose, we introduce the dimensionless integration variable
$y=2ak_{\bot}$ and rewrite the first formula in Eq.~(\ref{eq50})
in the form
\begin{equation}
\Delta_T^{\!(2)}{\cal F}_{\rm TM}^{(l=0)}(a,T)=
-\frac{8(k_BT)^2}{\pi^2\hbar ca(\pi\alpha+2\tilde{v}_F)}
Y_{\rm TM}(a,T),
\label{eq51}
\end{equation}
\noindent
where
\begin{equation}
Y_{\rm TM}(a,T)=\int_{0}^{\infty}\!\!\!dy
\frac{\rho_{\rm TM}^2}{e^{y}-\rho_{\rm TM}^2}
\int_{0}^{1}\!\!\!dx\ln\left[1+e^{-Ay\sqrt{x(1-x)}}\right],
\label{eq52}
\end{equation}
\noindent
and, according to Eq.~(\ref{eq33}),
\begin{equation}
A=\frac{\hbar v_F}{2ak_BT}=\frac{T_{\rm eff}^{(g)}}{T}\gg 1.
\label{eq53}
\end{equation}

Using the series expansions and calculating the integral with
respect to $y$, Eq.~(\ref{eq52}) is rewritten as
\begin{eqnarray}
&&
Y_{\rm TM}(a,T)=\sum_{n=1}^{\infty}\rho_{\rm TM}^{2n}
\sum_{k=1}^{\infty}\frac{(-1)^{k+1}}{k}
\int_{0}^{1}\!\! dx
\int_{0}^{\infty}\!\!\!dy
e^{-ny-Aky\sqrt{x(1-x)}}
\label{eq54} \\
&&~~
=\sum_{n=1}^{\infty}\rho_{\rm TM}^{2n}
\sum_{k=1}^{\infty}\frac{(-1)^{k+1}}{k}
\int_{0}^{1}\!\! \frac{dx}{n+Ak\sqrt{x(1-x)}}.
\nonumber
\end{eqnarray}

Taking into account that $A\to\infty$ when $T\to 0$,
one can write
\begin{equation}
Y_{\rm TM}(a,T)=\frac{1}{A}\sum_{n=1}^{\infty}\rho_{\rm TM}^{2n}
\sum_{k=1}^{\infty}\frac{(-1)^{k+1}}{k^2}
\int_{0}^{1}\!\! \frac{dx}{\sqrt{x(1-x)}}
=\frac{1}{A}\,\frac{\rho_{\rm TM}^2}{1-\rho_{\rm TM}^2}
\,\frac{\pi^3}{12}.
\label{eq55}
\end{equation}
\noindent
Note that an exact calculation of the integral in Eq.~(\ref{eq54})
leads to the same result under the condition (\ref{eq53}).

Substituting Eqs.~(\ref{eq36}) and (\ref{eq55}) in
Eq.~(\ref{eq51}), one arrives at
\begin{equation}
\Delta_T^{\!(2)}{\cal F}_{\rm TM}^{(l=0)}(a,T)=
-\frac{(k_BT)^3}{(\hbar v_F)^2}
\frac{\pi^3\alpha^2}{3(\pi\alpha+2\tilde{v}_F)
(\pi\alpha+\tilde{v}_F)}.
\label{eq56}
\end{equation}
\noindent
{}From the comparison of Eqs.~(\ref{eq44}) (the first line) and
(\ref{eq56}), it is seen that the thermal corrections
$\Delta_T^{\!(1)}{\cal F}_{\rm TM}$ and
$\Delta_T^{\!(2)}{\cal F}_{\rm TM}^{(l=0)}$
have the same sign, corresponding to attraction, and their
absolute values are of the same order of magnitude.

Now we return to the contribution of the TE mode to
$\Delta_T^{\!(2)}{\cal F}^{(l=0)}$ defined in the second formula
of Eq.~(\ref{eq50}). We rewrite this formula in terms of the variable
$y$ using Eq.~(\ref{eq36}) and neglecting by
$\pi\alpha\tilde{v}_F$
in comparison with 2. The result is
\begin{equation}
\Delta_T^{\!(2)}{\cal F}_{\rm TE}^{(l=0)}(a,T)=
\frac{k_BT\alpha^2\tilde{v}_F^2}{a^2}Y_{\rm TE}(a,T),
\label{eq57}
\end{equation}
\noindent
where
\begin{equation}
Y_{\rm TE}(a,T)=\int_{0}^{\infty}ydye^{-y}
\int_{0}^{1}dx
\frac{\sqrt{x(1-x)}}{e^{Ay\sqrt{x(1-x)}}+1}.
\label{eq58}
\end{equation}
\noindent
By expanding in the power series and integrating with respect to
$y$, one finds
\begin{eqnarray}
&&
Y_{\rm TE}(a,T)=\sum_{k=1}^{\infty}(-1)^{k+1}
\int_{0}^{1}dx
\sqrt{x(1-x)}
\int_{0}^{\infty}ydye^{-y-Aky\sqrt{x(1-x)}}
\label{eq59} \\
&&~~~~~~~~
=\sum_{k=1}^{\infty}(-1)^{k+1}
\int_{0}^{1}dx
\frac{\sqrt{x(1-x)}}{[1+Ak\sqrt{x(1-x)}]^2}.
\nonumber
\end{eqnarray}

Taking into account Eq.~(\ref{eq53}), we can neglect by the unity
in the denominator and write
\begin{equation}
Y_{\rm TE}(a,T)=\frac{1}{A^2}\sum_{k=1}^{\infty}
\frac{(-1)^{k+1}}{k^2}
\int_{0}^{1}\frac{dx}{\sqrt{x(1-x)}}.
\label{eq60}
\end{equation}
\noindent
Integrating with respect to $x$ and summing in $k$, one arrives
at
\begin{equation}
Y_{\rm TE}(a,T)=\frac{\pi^3}{12A^2}=\frac{\pi^3}{12}
\left(\frac{2ak_BT}{\hbar v_F}\right)^2.
\label{eq61}
\end{equation}

Substituting this result in Eq.~(\ref{eq57}), we find
\begin{equation}
\Delta_T^{\!(2)}{\cal F}_{\rm TE}^{(l=0)}(a,T)=
\frac{(k_BT)^3}{(\hbar c)^2}\,\frac{\pi^3\alpha^2}{3}.
\label{eq62}
\end{equation}

Thus, the TE contribution to the thermal correction is again
of opposite sign to the TM contribution (\ref{eq56}).
{}From Eqs.~(\ref{eq56}) and (\ref{eq62}), we have
\begin{equation}
\frac{\Delta_T^{\!(2)}{\cal F}_{\rm TE}^{(l=0)}}{|\Delta_T^{\!(2)}
{\cal F}_{\rm TM}^{(l=0)}|}=
\tilde{v}_F^2(\alpha\pi+2\tilde{v}_F)(\alpha\pi+\tilde{v}_F)
\ll 1,
\label{eq63}
\end{equation}
\noindent
i.e., the TE contribution to this part of the thermal
correction is again much smaller than the TM contribution.

\subsection{Contribution of all terms with nonzero
Matsubara frequencies}

Now we consider the contribution to the thermal correction
$\Delta_T^{(2)}{\cal F}$ due to all nonzero Matsubara frequencies.
 We again represent the quantity of our interest as the sum of
TM and TE modes
\begin{equation}
\Delta_T^{(2)}{\cal F}^{(l\geq 1)}(a,T)=
\Delta_T^{(2)}{\cal F}_{\rm TM}^{(l\geq 1)}(a,T)+
\Delta_T^{(2)}{\cal F}_{\rm TE}^{(l\geq 1)}(a,T),
\label{eq64}
\end{equation}
\noindent
where according to Eq.~(\ref{eq15})
\begin{eqnarray}
&&
\Delta_T^{\!(2)}{\cal F}_{\rm TM}^{(l\geq 1)}(a,T)=-\frac{2k_BT}{\pi}
\sum_{l=1}^{\infty}\int_{0}^{\infty}\!\!k_{\bot}dk_{\bot}
\nonumber\\
&&~
\times\frac{\tilde{q}_l}{\pi\alpha q_l+2\tilde{q}_l}\,
\frac{{r_{\rm TM}^{(0)}\!}^2(i\xi_l,k_{\bot})}{e^{2q_la}-
{r_{\rm TM}^{(0)}\!}^2(i\xi_l,k_{\bot})}\,
\frac{\Delta_T\Pi_{00}(i\xi_l,k_{\bot})}{\Pi_{00}^{(0)}(i\xi_l,k_{\bot})},
\nonumber\\
&&
\Delta_T^{\!(2)}{\cal F}_{\rm TE}^{(l\geq 1)}(a,T)=-\frac{2k_BT}{\pi}
\sum_{l=1}^{\infty}\int_{0}^{\infty}\!\!k_{\bot}dk_{\bot}
\label{eq65}\\
&&~
\times
\frac{q_l}{\pi\alpha\tilde{q}_l+2q_l}\,
\frac{{r_{\rm TE}^{(0)}}^2(i\xi_l,k_{\bot})}{e^{2q_la}-
{r_{\rm TE}^{(0)}}^2(i\xi_l,k_{\bot})}\,
\frac{\Delta_T\Pi(i\xi_l,k_{\bot})}{\Pi^{(0)}(i\xi_l,k_{\bot})}.
\nonumber
\end{eqnarray}

As shown in Appendix~A, at low temperature, satisfying
Eq.~(\ref{eq33}),
and for $l\geq 1$, Eq.~(\ref{eq9}) leads to
\begin{eqnarray}
&&
\Delta_T\Pi_{00}(i\xi_l,k_{\bot})=
\frac{6\zeta(3)\alpha\hbar k_{\bot}^2}{\tilde{q}_lB_l^3},
\label{eq66} \\
&&
\Delta_T\Pi(i\xi_l,k_{\bot})=
\frac{12\zeta(3)\alpha\hbar k_{\bot}^2\tilde{q}_l}{B_l^3}
\left(\frac{3\xi_l^2}{2c^2\tilde{q}_l^2}-1\right).
\nonumber
\end{eqnarray}

We begin with the
contribution of the TM mode given by the first formula
in Eq.~(\ref{eq65}). {}From Eqs.~(\ref{eq66}) and (\ref{eq6}),
using the variable $y$ defined in Eq.~(\ref{eq22}) and
the effective
temperature defined in Eq.~(\ref{eq23}), one obtains
\begin{equation}
\frac{\Delta_T\Pi_{00}}{\Pi_{00}^{(0)}}=
\frac{48\zeta(3)}{\pi\tilde{g}_l^3}
\left(\frac{T}{T_{\rm eff}}\right)^3,
\label{eq67}
\end{equation}
\noindent
where $\tilde{g}_l$ is defined in Eq.~(\ref{eq27}).

Now we rewrite the first formula in Eq.~(\ref{eq65}) in terms of
the variable $y$ and use Eq.~(\ref{eq67}). The result is
\begin{equation}
\Delta_T^{\!(2)}{\cal F}_{\rm TM}^{(l\geq 1)}(a,T)=
-\frac{24\zeta(3)k_BT}{\pi^2a^2}
\left(\frac{T}{T_{\rm eff}}\right)^3\,X_{\rm TM}(a,T),
\label{eq68}
\end{equation}
\noindent
where
\begin{equation}
X_{\rm TM}(a,T)=\sum_{l=1}^{\infty}\int_{\tau l}^{\infty}\!\!
\frac{ydy}{\tilde{g}_l^2}\,\frac{1}{\pi\alpha y+2\tilde{g}_l}\,
\frac{{r_{\rm TM}^{(0)}}^2}{e^{y}-{r_{\rm TM}^{(0)}}^2}
\label{eq69}
\end{equation}
\noindent
and $r_{\rm TM}^{(0)}\equiv r_{\rm TM}^{(0)}(y,\zeta_l)$ is
defined in Eq.~(\ref{eq26}).

Expanding under the integral in powers of
${r_{\rm TM}^{(0)}}^2e^{-y}$, we can rewrite Eq.~(\ref{eq69}) as
\begin{equation}
X_{\rm TM}(a,T)=\sum_{l=1}^{\infty}
\sum_{n=1}^{\infty}\int_{\tau l}^{\infty}\!\!
\frac{ydy}{\tilde{g}_l^2}\,
\frac{{r_{\rm TM}^{(0)}}^{2n}e^{-ny}}{\pi\alpha y+2\tilde{g}_l}.
\label{eq70}
\end{equation}

It is convenient to use Eq.~(\ref{eq26}) for
${r_{\rm TM}^{(0)}}$ and introduce a new integration
variable $x=y/(\tau l)$. Then, Eq.~(\ref{eq70}) takes the form
\begin{equation}
X_{\rm TM}(a,T)=\sum_{l=1}^{\infty}
\sum_{n=1}^{\infty}\frac{e^{-nl\tau}}{\tau l}\,
Q_n(\tau l),
\label{eq71}
\end{equation}
\noindent
where
\begin{equation}
Q_n(\tau l)=\int_{1}^{\infty}\!\!\frac{xdx}{(\tilde{v}_F^2x^2+1)}
\left(\frac{\alpha\pi x}{\alpha\pi x+2\sqrt{\tilde{v}_F^2x^2+1}}
\right)^{2n}
\frac{e^{-nl\tau(x-1)}}{\alpha\pi x+
2\sqrt{\tilde{v}_F^2x^2+1}}.
\label{eq72}
\end{equation}

We are interested in finding the leading term of the quantity
$Q_n(\tau l)$ when $\tau$ goes to zero. This is given by
\begin{equation}
Q_n(0)=\int_{1}^{\infty}\!\!\frac{xdx}{(\tilde{v}_F^2x^2+1)}
\left(\frac{\alpha\pi x}{\alpha\pi x+2\sqrt{\tilde{v}_F^2x^2+1}}
\right)^{\!2n}
\frac{1}{\alpha\pi x+2\sqrt{\tilde{v}_F^2x^2+1}}.
\label{eq73}
\end{equation}
\noindent
The correction terms to Eq.~(\ref{eq73}) due to nonzero $\tau$
go to zero when $\tau$ vanishes. This is seen from the following:
\begin{equation}
|Q_n(\tau l)-Q_n(0)|<\frac{1}{\tilde{v}_F^2}\left(
\frac{\alpha\pi}{\alpha\pi+2\tilde{v}_F}\right)^{\!2n}
\frac{1}{\alpha\pi+2\tilde{v}_F}
\int_{1}^{\infty}\!\frac{dx}{x^2}\left[1-e^{-nl\tau(x-1)}
\right]
\label{eq74}
\end{equation}
\noindent
and from the fact that the integral entering the right-hand
side of this equation goes to zero when $\tau\to 0$:
\begin{equation}
-nl\tau\,e^{nl\tau}\,{\rm Ei}(-nl\tau)\to 0.
\label{eq74a}
\end{equation}

We find the leading term of the quantity (\ref{eq71}) in the
limiting case $\tau\to 0$ by substituting $Q_n(0)$ in place of
$Q_n(\tau l)$. In so doing, the summation in $l$ is performed
according to
\begin{equation}
\sum_{l=1}^{\infty}\frac{e^{-nl\tau}}{l}=-\ln(1-e^{-n\tau})
\approx
-\ln(\tau n)=-\ln\tau -\ln n,
\label{eq75}
\end{equation}
\noindent
where for obtaining the leading term one should keep only
the first contribution on the right-hand side. Performing also
the trivial summation in $n$, one obtains
\begin{equation}
X_{\rm TM}(a,T)=-\frac{\ln\tau}{\tau}C_{\rm TM},
\label{eq76}
\end{equation}
\noindent
where the constant $C_{\rm TM}$ is given by
\begin{equation}
C_{\rm TM}=\frac{\alpha^2\pi^2}{4}\!\int_{1}^{\infty}
\!\!\!\!\frac{x^3dx}{(\tilde{v}_F^2x^2+1)^{3/2}}
\frac{1}{(\alpha\pi x+\sqrt{\tilde{v}_F^2x^2+1})
(\alpha\pi x+2\sqrt{\tilde{v}_F^2x^2+1})}.
\label{eq77}
\end{equation}
\noindent
The numerical integration in Eq.~(\ref{eq77}) results in
$C_{\rm TM}\approx 1.3\times 10^4$.

Substituting Eq.~(\ref{eq76}) in Eq.~(\ref{eq68}), we
arrive at
\begin{equation}
\Delta_T^{\!(2)}{\cal F}_{\rm TM}^{(l\geq 1)}(a,T)=
\frac{(k_BT)^3}{(\hbar c)^2}
\ln\left(\frac{ak_BT}{\hbar c}\right)
\frac{48\zeta(3)}{\pi^3}\,C_{\rm TM}.
\label{eq78}
\end{equation}
\noindent
Note that we have omitted the factor $4\pi$ under the
logarithm because it contributes to the next after the
leading term of order of $T^3$.

Now we continue with the contribution of the TE mode to
the thermal correction
$\Delta_T^{\!(2)}{\cal F}^{(l\geq 1)}$.
{}From the second formulas of Eqs.~(\ref{eq66}) and
(\ref{eq6}), one obtains
\begin{equation}
\frac{\Delta_T\Pi}{\Pi^{(0)}}=
\frac{96\zeta(3)}{\pi\tilde{g}_l^3}
\left(\frac{T}{T_{\rm eff}}\right)^3
\left(\frac{3\tau^2l^2}{2\tilde{g}_l^2}-1\right).
\label{eq79}
\end{equation}

Substituting this results in the second formula of
Eq.~(\ref{eq65}), we find
\begin{equation}
\Delta_T^{\!(2)}{\cal F}_{\rm TE}^{(l\geq 1)}(a,T)=
-\frac{48\zeta(3)k_BT}{\pi^2a^2}
\left(\frac{T}{T_{\rm eff}}\right)^3\,X_{\rm TE}(a,T),
\label{eq80}
\end{equation}
\noindent
where
\begin{equation}
X_{\rm TE}(a,T)=\sum_{l=1}^{\infty}\int_{\tau l}^{\infty}\!\!
\frac{y^2dy}{\tilde{g}_l^3}\,\frac{1}{\pi\alpha\tilde{g}_l+2y}
\frac{{r_{\rm TE}^{(0)}}^2}{e^{y}-{r_{\rm TE}^{(0)}}^2}
\left(\frac{3\tau^2l^2}{2\tilde{g}_l^2}-1\right)
\label{eq81}
\end{equation}
\noindent
and $r_{\rm TE}^{(0)}\equiv r_{\rm TE}^{(0)}(y,\zeta_l)$ is
defined in Eq.~(\ref{eq26}).

Similarly to the case of TM mode, we
expand  in powers of
${r_{\rm TE}^{(0)}}^2e^{-y}$ and obtain
\begin{equation}
X_{\rm TE}(a,T)=\sum_{l=1}^{\infty}
\sum_{n=1}^{\infty}\int_{\tau l}^{\infty}\!\!
\frac{y^2dy}{\tilde{g}_l^3}\,
\frac{{r_{\rm TE}^{(0)}}^{2n}e^{-ny}}{\pi\alpha\tilde{g}_l+2y}
\left(\frac{3\tau^2l^2}{2\tilde{g}_l^2}-1\right).
\label{eq82}
\end{equation}

Introducing the integration variable $x=y/(\tau l)$
and using Eq.~(\ref{eq26}) for
${r_{\rm TE}^{(0)}}$, one arrives at
\begin{equation}
X_{\rm TE}(a,T)=\sum_{l=1}^{\infty}
\sum_{n=1}^{\infty}\frac{e^{-nl\tau}}{\tau l}\,
G_n(\tau l),
\label{eq83}
\end{equation}
\noindent
where
\begin{eqnarray}
&&
G_n(\tau l)=\int_{1}^{\infty}\!\!\frac{x^2dx}{(\tilde{v}_F^2x^2+1)^{3/2}}
\left(\frac{\alpha\pi\sqrt{\tilde{v}_F^2x^2+1}}{\alpha\pi
\sqrt{\tilde{v}_F^2x^2+1}+2x}
\right)^{2n}
\nonumber \\
&&~~
\times
\frac{e^{-nl\tau(x-1)}}{\alpha\pi\sqrt{\tilde{v}_F^2x^2+1}+2x}
\left[\frac{3}{2(\tilde{v}_F^2x^2+1)}-1\right].
\label{eq84}
\end{eqnarray}

Similarly to the case of the TM mode,
the leading term of Eq.~(\ref{eq84}) at small $\tau$
is given by the value of $G_n$ at $\tau=0$.
Substituting $G_n(0)$ in place of $G_n(\tau l)$ in
Eq.~(\ref{eq83}) and performing the summations in $l$ and $n$,
we obtain
\begin{equation}
X_{\rm TE}(a,T)=-\frac{\ln\tau}{\tau}C_{\rm TE},
\label{eq85}
\end{equation}
\noindent
where
\begin{eqnarray}
&&
C_{\rm TE}=\frac{\alpha^2\pi^2}{4}\!\int_{1}^{\infty}
\!\!\!\!\frac{dx}{\sqrt{\tilde{v}_F^2x^2+1}(\alpha\pi
\sqrt{\tilde{v}_F^2x^2+1}+x)}
\nonumber \\
&&~~~
\times
\frac{1}{\alpha\pi\sqrt{\tilde{v}_F^2x^2+1}+2x}
\left[\frac{3}{2(\tilde{v}_F^2x^2+1)}-1\right].
\label{eq86}
\end{eqnarray}
\noindent
According to the results of
the numerical integration,
$C_{\rm TE}\approx 1.1\times 10^{-4}$.

Finally, the substitution of Eq.~(\ref{eq85}) in
Eq.~(\ref{eq80}) leads to
\begin{equation}
\Delta_T^{\!(2)}{\cal F}_{\rm TE}^{(l\geq 1)}(a,T)=
\frac{(k_BT)^3}{(\hbar c)^2}
\ln\left(\frac{ak_BT}{\hbar c}\right)
\frac{96\zeta(3)}{\pi^3}\,C_{\rm TE}.
\label{eq87}
\end{equation}

It is seen that
\begin{equation}
\frac{\Delta_T^{\!(2)}
{\cal F}_{\rm TE}^{(l\geq 1)}(a,T)}{\Delta_T^{\!(2)}
{\cal F}_{\rm TM}^{(l\geq 1)}(a,T)}=
2\frac{C_{\rm TE}}{C_{\rm TM}}\approx 1.7\times 10^{-8},
\label{eq88}
\end{equation}
\noindent
i.e., the contribution of the TE mode is again negligibly
small as compared to the contribution of the TM mode.

\section{Low-temperature behavior of the Casimir free energy,
{\protect \\}entropy and pressure for two graphene sheets}

According to the above results, at low temperatures,
satisfying the condition (\ref{eq33}), the leading term of
the thermal correction to the Casimir energy is given by
Eq.~(\ref{eq78}). This term is determined by an explicit
dependence of the polarization tensor on temperature as a
parameter and originates from all contributions
 to the Lifshitz formula with nonzero Matsubara
frequencies. We have shown also that at low temperature the
magnitudes of all other parts of the thermal correction to the
Casimir energy (determined by the zero-temperature polarization
tensor calculated at the Matsubara frequencies and by the thermal
correction to it calculated at zero frequency) are of the
next-to-leading order. In all cases the TM mode gives the major
contribution to the result.

Thus, in accordance with Eq.~(\ref{eq20}), the Casimir free energy
at low temperature can be written as
\begin{equation}
{\cal F}(a,T)=E(a)+
\frac{(k_BT)^3}{(\hbar c)^2}
\ln\left(\frac{ak_BT}{\hbar c}\right)
\frac{48\zeta(3)}{\pi^3}\,C_{\rm TM},
\label{eq89}
\end{equation}
\noindent
where the Casimir energy $E(a)$ at $T=0$ is defined in
Eq.~(\ref{eq16}) and
the coefficient $C_{\rm TM}$ is given in Eq.~(\ref{eq77}).
As is seen from Eq.~(\ref{eq89}), the thermal correction has
the same (negative) sign as the Casimir energy, which
corresponds to an attraction.

Now we are in a position to calculate the Casimir entropy
of two graphene sheets in the limit of low temperatures.
This is of interest in connection with the problems arising
for two Casimir plates made of metals or dielectrics
(see Sec.~I). {}From Eq.~(\ref{eq89}) we obtain the
leading contribution to the entropy in the form
\begin{equation}
S(a,T)=-\frac{\partial{\cal F}(a,T)}{\partial T}
=
-\left(\frac{k_BT}{\hbar c}\right)^2
\ln\left(\frac{ak_BT}{\hbar c}\right)
\frac{144\zeta(3)k_B}{\pi^3}\,C_{\rm TM}.
\label{eq90}
\end{equation}

As is seen in Eq.~(\ref{eq90}), the Casimir entropy of two
graphene sheets is positive. With decreasing $T$ we have
\begin{equation}
{\displaystyle{\lim_{T\to 0}}} S(a,T)=0,
\label{eq91}
\end{equation}
\noindent
which means that
the Nernst heat theorem is satisfied.
Thus, the Lifshitz theory of the Casimir interaction between two
graphene sheets is in agreement with thermodynamics in spite of
the fact that some of its properties (specifically, large thermal
effect at short separations and almost zero contribution of the
TE mode)  are reminiscent of that of metals described by the Drude
model. It is not surprising, however, that for graphene the
theory turns out to be thermodynamically consistent.
The reason is that the reflection coefficients (\ref{eq2}) are
expressed via the polarization tensor (\ref{eq5}), or
equivalent nonlocal dielectric permittivities (\ref{eq4}),
which are calculated starting from the first principles of
quantum electrodynamics at nonzero temperature.

As opposed to next-to-leading-order contributions to the
thermal correction presented in Eqs.~(\ref{eq44}), (\ref{eq56}),
and (\ref{eq62}), which are independent on the separation between
graphene sheets,
the leading term (\ref{eq87}) depends on $a$.
This makes possible to obtain an asymptotic expression
for the Casimir pressure between two graphene sheets at
vanishing temperature. {}From Eq.~(\ref{eq89}) one obtains
\begin{equation}
P(a,T)=-\frac{\partial{\cal F}(a,T)}{\partial a}=P_0(a)-
\frac{(k_BT)^3}{(\hbar c)^2}
\,
\frac{48\zeta(3)}{\pi^3}\,\frac{C_{\rm TM}}{a}.
\label{eq92}
\end{equation}
\noindent
As is seen from this equation, at low temperature the
thermal correction to the Casimir pressure is inversely
proportional to the separation distance.

At the end of this section, we characterize the experimental
situation. By now there is only one experiment on measuring
the Casimir interaction between a Au-coated sphere and a
graphene-coated substrate \cite{66}. Using the Lifshitz
theory with reflection coefficients expressed via the
polarization tensor of graphene and dielectric permittivity
of substrate, the measurement data were found to be in  very good
agreement with the calculation results \cite{67}.
Thus, as it was already demonstrated for metals and
dielectrics, the measurement results for graphene again
 confirm the thermodynamically consistent theory.

\section{Conclusions and discussion}

In the foregoing, we have investigated the low-temperature
behavior of the Casimir free energy between two graphene
sheets. This was done in the framework of a fundamental
theory using the polarization tensor in (2+1)-dimensional
space-time defined over the entire plane of complex frequency.
The thermal correction to the Casimir energy was separated in
two parts: the first one determined by the polarization
tensor at zero temperature calculated at the discrete
Matsubara frequencies, and the second one determined by the
thermal correction to the polarization tensor. The second
part of the thermal correction was subdivided into the
contributions of the zero-frequency term of the Lifshitz formula
and all terms with nonzero Matsubara frequencies.

Using the analytic asymptotic expansions in powers of small
parameters, we have shown that at all temperatures, which are
much smaller than the effective temperature of graphene
defined in Eq.~(\ref{eq33}), the leading terms in the first part
of the thermal correction to the Casimir energy and in the
zero-frequency contribution to the second part behave as
$\sim T^3$. The leading term in the second part of the thermal
correction, originating from all nonzero Matsubara frequencies,
behaves as $\sim T^3\ln{T}$. Thus, it determines the temperature
dependence   of the Casimir free energy at low temperature.

Note that at higher temperatures the situation is quite
different. Thus, at $T=300\,$K the dominant contributions to
the thermal effect are given by the first part of the thermal
correction and by the zero-frequency term in the second part
of the thermal correction \cite{33}. These are the contributions
which, according to our results, are of the next-to-leading order
at low temperatures. In so doing, at both low and room
temperatures the contributions of the TE mode to all the results
are much smaller in magnitude than the respective contributions of
the TM mode and can be neglected.

The obtained analytic expression for the Casimir free energy of
two graphene sheets at low temperature was used to calculate the
low-temperature behavior of the Casimir entropy and pressure.
It was shown that the Casimir entropy behaves as
$\sim T^2\ln{T}$ and, thus, goes to zero when the temperature
vanishes in accordance with the Nernst heat theorem.
The thermal correction to the Casimir pressure between two
graphene sheets behaves at low temperature as $\sim T^3/a$,
i.e., slowly increases with decreasing $a$. This should be
compared with the thermal correction to the Casimir pressure
between Drude metals, which behaves at low $T$ as
$P_D\sim T/a^3$ \cite{1,38}.

The obtained results shed new light on the problem of
theoretical  description of free charge carriers in the
Casimir physics. This problem at the moment is not of only
academic character. The point is that in the difference force
measurements \cite{49,50,51,52,53}, as well as for the Casimir
free energy and pressure of thin metallic films \cite{75,76,77},
the predictions of the Drude and plasma model approaches are
recently shown to differ by up to a factor
of several thousands. As we have shown here, an
experimentally consistent theory for graphene, using the polarization tensor
in (2+1)-dimensional space-time, is in agreement with the
requirements of thermodynamics similar to experimentally
consistent approaches for
metallic and dielectric test bodies. This confirms that the
thermodynamic test is important for future resolution of the
problem of relaxation properties of free charge carriers in the
Casimir physics.

%%%%%%%%%%%%%%%%%%%%%%%%%%%%%%%%%%%%%%%%%%%%%%%%%%%%%%%%%%
\section*{Acknowledgments}

The authors of this work acknowledge CNPq (Brazil) for
 partial financial support (the Grants 307596/2015--0 and
 308150/2015--5).
 They are also indebted to M.~Bordag for helpful discussions,
for reading the manuscript, and for useful corrections and
suggestions.
The work of V.M.M.~was partially supported by the Russian Government
Program of Competitive Growth of Kazan Federal University.
G.L.K.\ and V.M.M.\ are grateful to the Department
of Physics of the Federal University of
Para\'{\i}ba (Jo\~{a}o Pessoa, Brazil) for kind hospitality.

%%%%%%%%%%%%%%%%%%%%%%%%%%%%%%%%%%
\appendix
\section{}

Here, we derive the asymptotic expressions for the thermal
corrections, $\Delta_T\Pi_{00}$ and $\Delta_T\Pi$, valid
under the conditions (\ref{eq32}) and (\ref{eq33}) at all
nonzero Matsubara frequencies. We start from
$\Delta_T\Pi_{00}(i\xi_l.k_{\bot})$
presented in the first formula of
Eq.~(\ref{eq9}) and introduce the following notation for the
integral entering this formula
\begin{equation}
I_{00}=\int_{0}^{\infty}\frac{du}{e^{B_lu}+1}\left[1-
\frac{1}{\sqrt{2}}\sqrt{Z_{00}(u)}\right],
\label{A1}
\end{equation}
\noindent
where
\begin{equation}
Z_{00}=\sqrt{(1+u^2)^2-4
\frac{\tilde{v}_F^2k_{\bot}^2u^2}{\tilde{q}_l^2}}+1-u^2.
\label{A2}
\end{equation}

The major contribution to the integral (\ref{A1}) is given by
$u$ satisfying the condition $B_lu\sim 1$. Taking into account
that due to Eq.~({\ref{eq33})
\begin{equation}
B_l\equiv\frac{\hbar c\tilde{q}_l}{2k_BT}>
\frac{\hbar c\tilde{v}_Fk_{\bot}}{2k_BT}\sim
\frac{\hbar {v}_F}{4ak_BT}\gg 1,
\label{A3}
\end{equation}
\noindent
we conclude that the major contribution to the integral
(\ref{A1}) is given by $u\ll 1$.

When it is considered that
$\tilde{v}_F^2k_{\bot}^2/\tilde{q}_l^2<1$, in the region of $u$,
giving the major contribution to the integral (\ref{A1}),
the following condition is satisfied:
\begin{equation}
\frac{\tilde{v}_F^2k_{\bot}^2}{\tilde{q}_l^2}u^2\ll 1.
\label{A4}
\end{equation}

Now we expand the quantity $Z_{00}(u)$ in powers of the small
parameter (\ref{A4}) and obtain
\begin{equation}
Z_{00}(u)=1+u^2-
2\frac{\tilde{v}_F^2k_{\bot}^2}{\tilde{q}_l^2}
\frac{u^2}{1+u^2}+1-u^2.
\label{A5}
\end{equation}
\noindent
Then, neglecting $u^2$ as compared to unity in the
denominator, we find
\begin{equation}
Z_{00}(u)=2\left(1-
\frac{\tilde{v}_F^2k_{\bot}^2u^2}{\tilde{q}_l^2}
\right).
\label{A6}
\end{equation}

Substituting Eq.~(\ref{A6}) in Eq.~(\ref{A1}), expanding the
square root in the same small parameter (\ref{A4}), and
integrating with respect to $u$, one arrives at
\begin{equation}
I_{00}=\frac{\tilde{v}_F^2k_{\bot}^2}{2\tilde{q}_l^2}
\int_{0}^{\infty}\frac{u^2du}{e^{B_lu}+1}=
\frac{3\zeta(3)\tilde{v}_F^2k_{\bot}^2}{4\tilde{q}_l^2B_l^3}.
\label{A7}
\end{equation}

Finally, we substitute this equation in the first formula of
Eq.~(\ref{eq9}) and arrive at the first formula of
Eq.~(\ref{eq66}).

We emphasize that the above derivation is valid only for
nonzero Matsubara frequencies with $l\geq 1$. The point is
that for $l=0$ one has $\tilde{v}_Fk_{\bot}=\tilde{q}_0$
and, as a result, we have the term $(1-u^2)^2$ under the square
root in Eq.~(\ref{A2}). In this case, Eq.~(\ref{A5})
would be applicable only for $u<1$. This makes impossible
the substitution  of Eq.~(\ref{A6}) in Eq.~(\ref{A1}).

Note that if the higher-order terms in the  small parameter
(\ref{A4}) are taken into account in  Eq.~(\ref{A5}),
which leads to integrals of the form
\begin{equation}
I_{00}^{(k)}=\int_{0}^{\infty}\frac{u^{2k}du}{e^{B_lu}+1}=
\frac{C_k}{B_l^{2k+1}},
\label{A8}
\end{equation}
\noindent
where $k\geq2$. This would result in the higher-order
corrections $\sim(T/T_{\rm eff})^{2k+1}$ to $\Delta_T\Pi_{00}$,
which are omitted in our calculation.

Now we consider the thermal correction
$\Delta_T\Pi(i\xi_l,k_{\bot})$ presented in the second formula
of Eq.~(\ref{eq9}). The integral entering this formula can be
rewritten in the form
\begin{equation}
I=\int_{0}^{\infty}\frac{du}{e^{B_lu}+1}\left[
-\frac{\xi_l^2}{c^2}+
\frac{\tilde{q}_l^2}{\sqrt{2}}\sqrt{Z_{00}(u)}
Z(u)\right],
\label{A9}
\end{equation}
\noindent
where $Z_{00}$ is defined in Eq.~(\ref{A6}) and
\begin{equation}
Z(u)=1-\frac{\tilde{v}_F^2k_{\bot}^2}{\tilde{q}_l^2
\sqrt{(1+u^2)^2-4
\frac{\tilde{v}_F^2k_{\bot}^2u^2}{\tilde{q}_l^2}}}.
\label{A10}
\end{equation}

We again use the small parameter (\ref{A4}) and the expansion
(\ref{A6}) of $Z_{00}$ in powers of this parameter.
Expanding also the quantity $Z$ in powers of the same parameter,
we obtain
\begin{equation}
Z(u)=1-
\frac{\tilde{v}_F^2k_{\bot}^2}{\tilde{q}_l^2(1+u^2)}\left[1+
2\frac{\tilde{v}_F^2k_{\bot}^2u^2}{\tilde{q}_l^2(1+u^2)^2}
\right].
\label{A11}
\end{equation}
\noindent
Furthermore, expansion of Eq.~(\ref{A11}) up to the second order
in the small parameter $u$ results in
\begin{equation}
Z(u)=\frac{\xi_l^2}{c^2\tilde{q}_l^2}+
\frac{\tilde{v}_F^2k_{\bot}^2u^2}{\tilde{q}_l^2}
\left(1-
2\frac{\tilde{v}_F^2k_{\bot}^2}{\tilde{q}_l^2}\right).
\label{A12}
\end{equation}

Substituting Eqs.~(\ref{A6}) and (\ref{A12}) in Eq.~(\ref{A9}),
one arrives at
\begin{equation}
I=-\tilde{v}_F^2k_{\bot}^2\left(1-
\frac{3\xi_l^2}{2c^2\tilde{q}_l^2}\right)
\int_{0}^{\infty}\frac{u^2du}{e^{B_lu}+1}
=-\frac{3\zeta(3)\tilde{v}_F^2k_{\bot}^2}{2B_l^3}\,
\left(1-
\frac{3\xi_l^2}{2c^2\tilde{q}_l^2}\right).
\label{A13}
\end{equation}

Then, from the second formula of Eq.~(\ref{eq9}), we
arrive at the second formula of
Eq.~(\ref{eq66}).

Note that the expansion terms of $Z_{00}(u)$ and $Z(u)$, which
are of higher orders in $u$ than in Eqs.~(\ref{A6}) and
(\ref{A12}), lead to correction terms
of the same form, as in Eq.~(\ref{A8}), and to higher-order
corrections to $\Delta_T\Pi$ than the leading term
presented in Eq.~(\ref{eq66}).

%%%%%%%%%%%%%%%%%%%%%%%%%%%%%%%%%%%%%%%%%%%%%%%%%%%
%%%%%%%%%%%%%%%%%%%%%%%%%%%%

%%%%%%%%%%%%%%%%
\end{document}